\documentclass[a4paper,11pt]{article}

\usepackage{fullpage}
\usepackage[T1]{fontenc}

\usepackage{amsmath, amsthm, amssymb}
\usepackage{cite}

\usepackage{tikz}
\usetikzlibrary{shapes.geometric, arrows}

\usepackage{comment}

\allowdisplaybreaks

\usepackage{graphicx}
\usepackage{caption}
\usepackage{subcaption}
\usepackage[utf8]{inputenc}

\def\s{\sigma}

\def\p{\partial}

\newcommand{\be}{\begin{eqnarray}}
\newcommand{\ee}{\end{eqnarray}}

\def\F{{\cal F}}

\begin{document} 
	
\begin{titlepage}
	\thispagestyle{empty}
	\begin{flushright}
		
	\end{flushright}

	\vspace{35pt}
	
	\begin{center}
	    { \LARGE{\bf  de Sitter decay through goldstino evaporation}} 
		
		\vspace{50pt}
		
		{Fotis~Farakos$^{1,2}$, Alex~Kehagias$^{3}$ and Nikolaos~Liatsos$^{3}$}
		
		\vspace{25pt}

		$^1${\it  Dipartimento di Fisica ``Galileo Galilei''\\
		Universit\`a di Padova, Via Marzolo 8, 35131 Padova, Italy}
		
		\vspace{15pt}
		
	    $^2${\it   INFN, Sezione di Padova \\
		Via Marzolo 8, 35131 Padova, Italy}
		
		\vspace{15pt} 
		
        $^3${\it   Physics Division, National Technical University of Athens \\
        15780 Zografou Campus, Athens, Greece}
		
		\vspace{15pt}

		\vspace{40pt}
		
		{ABSTRACT} 
	\end{center}
	
We study  supergravity theories  with  supersymmetry intrinsically non-linearly realized supported only by four-form flux. We demonstrate that supersymmetry can be restored by the evaporation of the goldstino multiplet. Such procedure requires the existence of a super-extremal 2-brane, in accordance with  the WGC, and suggests an imminent decay in de Sitter supergravity, leaving the system to relax in supersymmetric AdS after the transition. In particular, the same decay mechanism also implies that the KKLT de Sitter vacua are short-lived. Our findings thus provide an effective realization of KPV-type of decays within 4D supergravity and in this way may reconcile the nilpotent superfield constructions of de Sitter with some aspects of the swampland program.

\vspace{10pt}

\bigskip

\end{titlepage}

\baselineskip 5.81 mm


\newpage

\tableofcontents

\section{Introduction }

The basic reasoning behind model-building within a 4D N=1 supergravity is that it can generically describe 
the low energy sector of specific string flux compactifications and may also capture some stringy phenomena, 
always within an effective field theory (EFT) approach. 
Within this mindset, 
cosmological models (late-time and inflationary) have been extensively studied in N=1 supergravity, 
and more recently the use of nilpotent superfields 
was put forward 
\cite{Antoniadis:2014oya,Ferrara:2014kva,DallAgata:2014qsj,Kallosh:2014hxa,Ferrara:2015gta,McDonough:2016der,
Cribiori:2016qif,Argurio:2017joe,Kallosh:2017wnt}. 
It was eventually recognized in \cite{Bergshoeff:2015tra} 
(earlier closely related work can be found in \cite{Lindstrom:1979kq,Farakos:2013ih,Dudas:2015eha}) 
that a pure de Sitter supergravity can be constructed which can describe the underlying sector of 
any late-time de Sitter cosmological phase. 
In particular such setup can be utilized to provide an effective description of the uplifting 
anti-D3-brane in the KKLT de Sitter construction of string theory \cite{Kachru:2003aw}, 
as was readily shown in \cite{Ferrara:2014kva} 
(the nilpotency being related to the intrinsically non-linear supersymmetry 
of the anti-D3-brane \cite{Kallosh:2014wsa,Bergshoeff:2015jxa,Garcia-Etxebarria:2015lif,Polchinski:2015bea,Akrami:2018ylq}). 
Constrained superfields can be further used to capture aspects 
of the open string sector \cite{Vercnocke:2016fbt,Aalsma:2017ulu,Kallosh:2016aep,GarciadelMoral:2017vnz,Cribiori:2019hod,Cribiori:2019bfx,Cribiori:2020bgt,Parameswaran:2020ukp}.

In the mean time, the concern that no  
de Sitter vacua can exist in string theory has been blatantly expressed in \cite{Danielsson:2018ztv,Obied:2018sgi}. 
This is formulated in terms of a swampland conjecture \cite{Obied:2018sgi} and its refinements 
\cite{Andriot:2018wzk,Garg:2018reu,Ooguri:2018wrx} according to which the low-energy effective potential should satisfy 
\begin{eqnarray}
\begin{aligned}
&|\nabla V|\geq c \frac{V}{M_{\rm Pl}} \, , \\
\mbox{or} \quad &
\mathrm{min}\Big(\nabla_i\nabla_j V\Big) \leq -c' \frac{V}{M_{\rm Pl}^2} \, , 
\end{aligned}
\end{eqnarray}
where $c,c'$ are numerical parameters of order ${\cal O}(1)$ and the derivatives refer to scalar field space. 
Therefore, according to this conjecture, metastable dS vacua are in the swampland and cannot emerge in a consistent quantum gravity theory whatever the latter is.
However, the Trans-Planckian Censorship Conjecture (TCC) \cite{Bedroya:2019tba}  is less restrictive and allows metastable dS vacua  \cite{Bedroya:2020rac} as long as they are short lived. 
It should be noted though that the TCC has been criticized in \cite{DKR} on the basis of the decoupling property of effective field theories. 
Nevertheless, there may exist a formulation of the TCC that still allows metastable dS vacua without clashing with  fundamental properties  of  effective field theories. 
We would like also to mention that, within the swampland program, 
the existence of long-lived non-supersymmetric AdS vacua has been also challenged in \cite{Ooguri:2016pdq}. 
Notably, 
the decay of non-SUSY AdS$_4$ flux vacua is tied to the existence of super-extremal 2-branes. 
Such non-SUSY AdS$_4$ vacua are also intrinsically related to the de Sitter vacua of string theory. 
Indeed, in KKLT for example one has to assume the possibility that a fully stable non-SUSY AdS could exist, 
even if our Universe is not described by such.

However, independent of the tachyonic instabilities, 
the story from the EFT perspective cannot yet be complete, 
because, for one, it is known that in a full string theory setup anti-branes can decay \cite{Kachru:2002gs,Gautason:2015tla}. 
Thus an instability even in non-tachyonic stringy de Sitter vacua may still be lurking around the corner 
(for a recent discussion see e.g. \cite{Farakos:2020idt}). 
In this work we aim exactly to fill this gap in the 4D N=1 supergravity EFT constructions, 
and we ask: what hidden decay channels could there be in the simplest de Sitter supergravity models? 
In particular we are interested in channels that describe the decay of the anti-D3-brane from a purely N=1 supergravity perspective 
and lead to the restoration of supersymmetry within the same supergravity model. 
Since the uplifting anti-D3-brane is generically described by a nilpotent chiral superfield, 
then a channel that describes the decay of the former in string theory has to lead to goldstino evaporation in the EFT. 
Indeed, 
supersymmetry cannot be restored in the presence of a goldstino, 
and since the non-linear supersymmetry here is intrinsic (bosonic DOF and fermionic don't match), 
the full goldstino has to vanish when supersymmetry gets restored. 
However, 
it is evident that such a transition between intrinsically non-linear supersymmetry 
and the supersymmetry restoration cannot take place in a continuous way in the same EFT 
because the goldstino couplings become huge near the supersymmetric point 
(even though the supersymmetric point per se can be consistent as we will see).

To overcome the above difficulties the first step 
is to utilize a nilpotent three-form chiral superfield to accommodate the goldstino \cite{Farakos:2016hly,Buchbinder:2017vnb}. 
In such a scenario the supersymmetry breaking order parameter is given by the four-form flux 
\be
f = \star dC_3 \, . 
\ee
Then, 
once we couple the $C_3$ to a membrane (2-brane) the four-form flux $\star dC_3$ can take discrete values and thus 
we can realize transitions via membrane nucleation \cite{BT1,BT2} between non-vanishing and vanishing $f$ in a discrete way 
(the former describing a state with intrinsically non-linear supersymmetry 
and the latter a state with supersymmetry being restored). 
In this way one can avoid the aforementioned strong coupling problem. 
Interestingly, 
we will see that it is exactly the existence of a super-extremal 2-brane charged under the goldstino three-form $C_3$ 
that allows such decay to take place, similarly to the situation in \cite{Ooguri:2016pdq}. 
In the simple examples we work out here the decay of de Sitter is imminent as the vacua 
are very short-lived once the super-extremal 2-brane decay channel is switched-on.

To summarize, we will examine here the decay of dS vacua 
where supersymmetry is non-linearly realized under the following conditions: 
1) There is a three-form chiral multiplet coupled to supergravity, 
2) supersymmetry is broken by the three-form multiplet with the latter satisfying a nilpotency condition, 
3) the theory has membranes (codimension-1 objects) and 
4) the Weak Gravity Conjecture \cite{ArkaniHamed:2006dz} holds 
and thus there are membranes with charge exceeding their tension. 
Under these conditions, we will see that such dS vacua where supersymmetry is broken are short-lived metastable dS spaces.

\section{Three-forms and supersymmetry breaking}

In this section we work in a globally supersymmetric setup and we describe the mechanism via which the 
goldstino multiplet can evaporate leading to supersymmetry restoration. 
There exist various ways to describe the goldstino, 
but it has been already shown in a series of papers \cite{DallAgata:2016syy,Cribiori:2016hdz,Cribiori:2017ngp} 
that we can generically describe the supersymmetry breaking sector of any system 
with a chiral superfield $X$ that satisfies the constraint $X^2=0$ so we will work directly with the nilpotent superfield formulation.

\subsection{Can the nilpotent superfield evaporate?}

Let us first consider the chiral superfield $X$ that can be written in chiral coordinates as \cite{Wess:1992cp} 
\be
\label{X-un}
X = A + \sqrt 2 \theta G + \theta^2 F \, . 
\ee 
Then we can impose the nilpotency constraint \cite{Rocek:1978nb,Casalbuoni:1988xh} 
\be
\label{X2zero}
X^2 = 0 \ \ \to \ \ \left\{ A^2, A G_\alpha, G^2 - 2 A F \right\} = 0 \, . 
\ee 
The equations \eqref{X2zero}, being quadratic, 
have two distinct algebraic solutions that are both completely compatible with the supersymmetry algebra. 
Which solution is picked by the system depends solely on whether or not the background value of the auxiliary field 
$F^{BG}$ has a vanishing value within the range of validity of the effective theory. 
The two solutions are: 
\begin{itemize}

\item[${\rm I})$] When $F^{BG} \ne 0$ supersymmetry is broken, 
the goldstino is described by the fermion $G_\alpha$, 
and the only fully consistent solution (consistent also with the supersymmetry algebra) is to have 
\be 
\label{AGGF}
A \equiv \frac{G^2}{2F} \, . 
\ee

\item[${\rm II})$] When there is at least one position with $F^{BG} = 0$ in the EFT then supersymmetry 
is preserved (at least by $X$), 
and the only fully consistent solution is to have that the full superfield vanishes identically, 
that is $X$ is anchored at 
\be
\label{XisZero}
X \equiv 0 \, .  
\ee

\end{itemize} 
Therefore the solution of the constraint relies entirely on the type of background we are studying. 
Notice that one could argue that the constraint \eqref{X2zero} can be solved for $A=0$ and $G_\alpha=0$ 
while $F \ne 0$, 
however such solution explicitly breaks supersymmetry therefore it should be discarded. 
Similarly one could try to use the solution \eqref{AGGF} for an EFT that passes from $F=0$, 
however at that point the solution becomes singular so it should be discarded. 
Moreover, 
the solution \eqref{AGGF} also allows for fluctuations of the $X$ superfield $\delta X$ 
as long as they satisfy 
\be
X^2=0 \ \ \to \ \  X \, \delta X =0 \, . 
\ee 
However, 
the solution \eqref{XisZero} completely neutralizes fluctuations because one has to impose 
\be
X = 0 \ \ \to \ \ \delta X = 0 \, .  
\ee

The simplest superspace Lagrangian that we can have for the nilpotent $X$ is given by \cite{Komargodski:2009rz} 
\be
\label{KS}
{\cal L}_X = \int d^4 \theta \, X \overline X 
+ \left( \int d^2 \theta \, f X + c.c. \right) \, , 
\ee
where $f$ is a positive real constant, and in component form using $A=G^2/2F$ and integrating out $F$ it leads to the Volkov--Akulov model \cite{Volkov:1973ix,Kuzenko:2010ef} 
for the goldstino 
\be
\label{KScomp}
{\cal L}_X= -f^2 + i \partial_m \overline G \overline \sigma^m G 
+ \frac{1}{4 f^2} \overline G^2 \partial^2 G^2 
-\frac{1}{16 f^6} G^2 \overline G^2 \partial^2 G^2 \partial^2 \overline G^2 \, . 
\ee
Such Lagrangian describes what we will call here a system with intrinsically non-linear supersymmetry 
because as we see even though the action is invariant under the supersymmetry transformations 
(this is manifest because we construct it from superspace), 
the fermionic and bosonic degrees of freedom do not match. 
Note that \eqref{KS} also defines a cut-off 
\be
\Lambda_{ cut-off} = \sqrt f \, . 
\ee 
We will use this as the cut-off of any effective theory where the goldstino enters 
(unless there is reason to have a lower cut-off) 
and we will also use it to define the tension of p-branes to 
be\footnote{A recent discussion on the membrane tension that is also relevant to our work can be found in \cite{Bedroya:2020rac}.} 
\be
T_{p+1} = (\Lambda_{cut-off})^{p+1} \, . 
\ee
Indeed, 
if we interpret the Volkov--Akulov action \eqref{KS} as a 3-brane \cite{Bandos:2015xnf} then we would assign to it a 
4D world-volume element $\sqrt{-g_4}$ which would consistently have a tension of the form 
\be
T_4 = (\Lambda_{cut-off})^4 = f^2 \, . 
\ee
This is in agreement with \eqref{KS}.

We can verify the validity of the solutions I) and II) of the nilpotency constraint 
when it is studied on different backgrounds also by studying the superspace equations of motion. 
One way to impose the nilpotency condition \eqref{X2zero} is with the use of a superspace Lagrange multiplier $\Psi$. 
To this end we can consider a Lagrangian with the unconstrained $X$ given by \eqref{X-un}, 
and we can simply add the Lagrange multiplier term with $\Psi$. 
We have 
\be
\label{LXPsi}
{\cal L}_{X-\Psi} = \int d^4 \theta \, X \overline X 
+ \left( \int d^2 \theta \, f X + c.c. \right) 
+ \left( \int d^2 \theta \, \Psi X^2 + c.c.  \right) \, . 
\ee
Then by integrating out the chiral superfield $\Psi$ we get the constraint \eqref{X2zero}, 
whereas by varying $X$ we get 
\be
\label{XX-EOM}
\frac14 \overline D^2 \overline X = f + 2 X \Psi \, . 
\ee
From the full superspace considerations we conclude again that 
combining \eqref{X2zero} and \eqref{XX-EOM} leads to two solutions depending 
on the value of $f$. 
When $f$ vanishes we have 
\be
f =0 \ \ \to \ \ X = 0 \, , 
\ee
and $\Psi$ remains undetermined, 
whereas when $f$ is non-vanishing it can sustain the supersymmetry breaking and we have 
\be
f \ne 0 \ \ \to \ \ A = \frac{G^2}{2F} \, , 
\ee
in which case one can also find the full non-trivial solution for the Lagrange multiplier $\Psi$.

\begin{figure}[ht]
\centering 
  \includegraphics[scale=0.5]{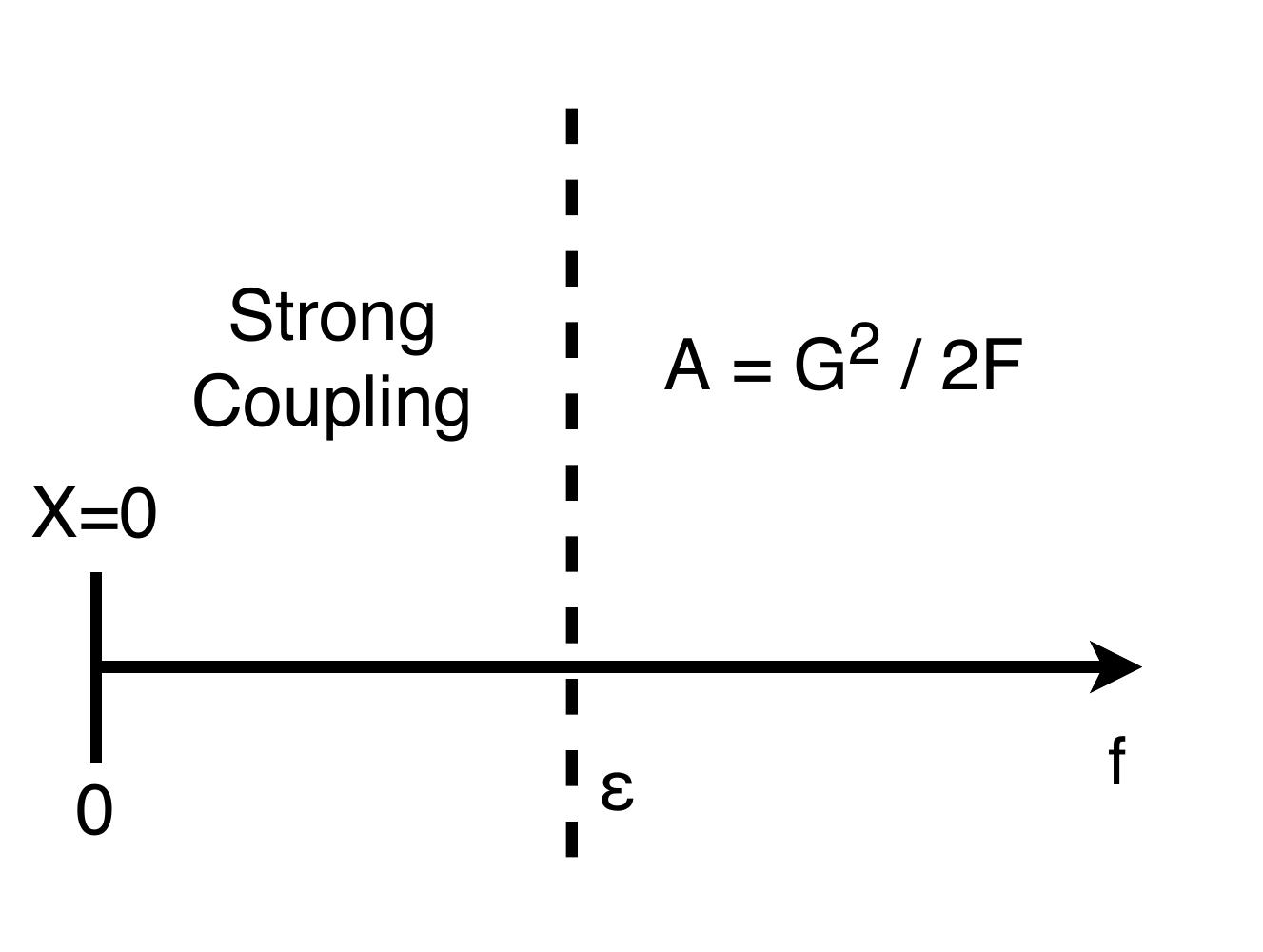} 
\caption{ 
{\it The regimes of validity of the different solutions to the nilpotency constraint. 
There is a region between $f=0$ and $f$ finite where the effective theory \eqref{KScomp} becomes strongly coupled.} 
\label{plot-REG}} 
\end{figure}

As a result we will see that if there can be systems with two distinct vacua 
that can be related by tunneling effects, 
in which case there will be a bubble of true vacuum $F=0$ growing inside 
the false vacuum $F \ne 0$, 
then the nilpotent goldstino multiplet can completely evaporate. 
Let us stress this point again: 
we are considering tunneling between the two distinct solutions \eqref{AGGF} and \eqref{XisZero} of \eqref{X2zero}, 
not between two different vacua of the same EFT arising from the solution \eqref{AGGF}. 
This is because \eqref{AGGF} and \eqref{XisZero} describe different EFTs with different regions of validity. 
Indeed, 
one should keep in mind that even though a configuration with $f \equiv 0$ can be self-consistent 
because the goldstino multiplet will evaporate, 
approaching such configuration in a continuous way may be problematic. 
In particular, 
if we approach the supersymmetric configuration in a continuous way by taking $f \to 0$, 
then, at some point, the $1 / \sqrt f$ will be large and the goldstino will enter a strong coupling regime. 
At that point the effective theory \eqref{KS} will not be reliable any more and we can not use it to draw any conclusions. 
In other words the regimes of validity of each of the two solutions \eqref{AGGF} and \eqref{XisZero} 
do not overlap and are separated by a parameter region where both solutions are not reliable. 
This situation is depicted in figure 1.

We have until now concluded that 
the transition from the intrinsically non-linear system \eqref{KS} to a supersymmetric background 
will certainly not be perturbative. 
We will see here one way this can happen: 
a spherical domain wall solution sourced by a membrane that interpolates between the aforementioned configurations. 
Note that the two vacua will have different Witten index \cite{Witten:1982df}, 
however, discontinuities in the Witten index are generically allowed so this is consistent. 
Such transition is depicted in figure 2. 
\begin{center}
	\begin{tikzpicture}[node distance=0cm]
	\tikzstyle{blobIN} = [circle, rounded corners, minimum width=3.5cm, minimum height=1cm, text centered, yshift=-0.35cm, 
	label=above:$F^{BG} \!\ne\! 0 \ \to \ A \!\equiv\! G^2 \!/ 2 F$, draw=black, fill=gray!01]
		\tikzstyle{blobOUT} = [rectangle, rounded corners, minimum width=6cm, minimum height=5cm, draw=black, fill=gray!07]
		\node (meta) [blobOUT] {}; 	
		\node (true) [blobIN] {$F^{BG} \!=\! 0 \ \to \ X\!\equiv\!0$}; 
	\end{tikzpicture}   
\end{center}
Figure\,2:\ {\it The two backgrounds/vacua are disconnected but there can be a bubble of the true vacuum growing 
within the meta-stable state. 
If the transition from $F^{BG} \ne 0$ to $F^{BG} = 0$ happens with a discrete step 
then the system will not pass through the strong coupling regime.} 
\setcounter{figure}{+2}

\subsection{Tunneling with 3-form multiplets}

A way to have transition between backgrounds with $f\ne0$ and $f=0$ is if $f$ is identified with the vacuum expectation value of the 
Hodge-dual of the field strength of a gauge three-form, 
i.e. if we have $f = \langle \star dC_3 \rangle$ \cite{Gates:1980ay,Binetruy:1996,Bandos:2010yy,Groh:2013}. 
We will see that there are nilpotent multiplets that can accommodate 
both the goldstino and the gauge three-form. 
The latter, by coupling to membranes, can jump and take different values. 
For the super-membrane setup in global supersymmetry we will follow mostly \cite{Bandos:2019qok}\footnote{In this subsection, we use the conventions of Appendix A of \cite{Bandos:2019qok} but the matrices $\s_{ab}$ and $\overline{\s}_{ab}$ are defined as in \cite{Wess:1992cp}.}.

A single three-form chiral superfield can be described with the use of a real prepotential $U$, 
and has the form 
\be
\label{X3form}
X = - \frac14 \overline D^2 U \, .  
\ee
The way the chiral projector acts on the real unconstrained $U$ implies that \eqref{X3form} is invariant under the superfield gauge transformation
\be
\label{UL}
U \to U + L \, , 
\ee
where $L$ is a real linear superfield satisfying the constraint $\overline D^2 L=0$. 
It is exactly this gauge invariance that signals that the multiplet \eqref{X3form} contains 
a gauge three-form in its F-term. 
This happens because the $\theta \s^m \overline \theta$ component of $U$, 
which is a real vector field, 
is varied under the gauge transformation \eqref{UL} by the $\theta \s^m \overline \theta$ component of the $L$ multiplet. 
However, 
the $\theta \s^m \overline \theta$ component of the $L$ multiplet is given by $(\star d B_2)_m$ for some real two-form $B_2$, 
which means that the $\theta \s^m \overline \theta$ component of the $U$ multiplet equals $(\star C_3)_m$ for some real three-form $C_3$ according to
\be
\label{Cm}
 - \frac18 \overline \sigma^{\dot\alpha  \alpha}_m [D_\alpha , \overline D_{\dot \alpha} ] U | =  (\star C_3)_m \equiv \frac{1}{3!} \epsilon_{mnpq} C^{npq} \equiv C_m \, . 
\ee
For us here, 
the only components that are relevant are the ones appearing in $X$. 
Following \cite{Groh:2013,Farakos:2016hly,Bandos:2019qok} we can define the component fields of $X$ as in \eqref{X-un} 
but now the F-term is composite and reads 
\be
\label{F3form}
F = -\frac{1}{4} {D}^2 X | =  \F + i \partial_m C^m = \F + i\star dC_3\, , 
\ee
where $\F \equiv \operatorname{Re} F$ is a real scalar that still serves as an auxiliary field, while the expression for the imaginary part of $F$ can be derived from equations \eqref{X3form} and \eqref{Cm} with use of the commutation relation $[D^2, \overline{D}^2]=-4i \overline {\s}^{m \hspace{0.05cm} \dot{\alpha} \alpha} \partial_m [D_\alpha , \overline D_{\dot \alpha}]$.

The bulk action for the nilpotent three-form superfield we will consider is described by a Lagrangian of the form 
\be
\label{nil-3form} 
{\cal L}_{X-\Psi} = \int d^4 \theta \, X \overline X 
+ \left( \int d^2 \theta \, \Psi X^2 + c.c.  \right) 
+ \mathcal{L}_{\text{bd}} \, , 
\ee
where $\Psi$ is a Lagrange multiplier chiral superfield enforcing the nilpotency condition \eqref{X2zero} and $\mathcal{L}_{\text{bd}}$ is a boundary term that ensures a consistent variation of the action with respect to the gauge three-form (see \cite{Groh:2013,Farakos:2017jme} for a review of this issue). We can derive this term by following a dualization procedure described  in \cite{Farakos:2017jme}. For this purpose, we consider the duality Lagrangian
\be
\label{Ldual}
\mathcal{L}_{\text{dual}}=\int d^4 \theta \, X \overline{X} +\left(\int d^2 \theta \, (\Psi X^2 + ZX) + c.c. \right) 
-\int d^4 \theta \, (Z+\overline{Z})U \, ,
\ee
where $U$ is a real superfield, $Z$ is a Lagrange multiplier chiral superfield and $X$ is now an ordinary chiral superfield. 
In order to find the formulation of \eqref{Ldual} in terms of a three-form multiplet, we need to integrate out the Lagrange multiplier $Z$ and the ordinary chiral superfield $X$. Varying \eqref{Ldual} with respect to $Z$ subject to the boundary condition $\delta Z|_{\text{bd}}=0$, we obtain
\be
\label{dZ}
X=-\frac{1}{4} \overline{D}^2 U \,,
\ee
which means that $X$ becomes a single three-form chiral superfield. Furthermore, the variation of \eqref{Ldual} with respect to $X$ subject to the boundary condition $\delta X|_{bd}=0$ leads to the following superspace equation of motion 
\be
\label{dX}
Z=\frac{1}{4} \overline{D}^2 \overline{X} - 2 \Psi X \,.
\ee
Rewriting \eqref{Ldual} as 
\be
\label{Ldual1}
\begin{aligned}
\mathcal{L}_{\text{dual}} = & \int d^4 \theta \, X \overline{X} + \left( \int d^2 \theta \, \Psi X^2 + c.c. \right) 
\\ & 
+  \left(\int d^2 \theta \, Z X + \frac{1}{8} \int d^2 \theta \overline{D}^2[(Z+\overline{Z})U] + c.c. 
\right) , 
\end{aligned}
\ee
and substituting equations \eqref{dZ} and \eqref{dX} into \eqref{Ldual1} we get the Lagrangian \eqref{nil-3form}, where $X$ is no longer an ordinary chiral superfield but a single three-form one given by \eqref{dZ} and the boundary term is given by
\be
\label{BD}
\mathcal{L}_{\text{bd}}=-\frac{1}{8}\left( \int d^2 \theta \, \overline{D}^2 - \int d^2 \overline{\theta} \, D^2 \right) \left[ \left( \frac{1}{4} \overline{D}^2 \overline{X} - 2 \Psi X \right) U \right] + c.c. 
\ee 
Now we wish to write \eqref{nil-3form} in component form. 
Let us first have a look at the gauge three-form sector in \eqref{nil-3form} which reads 
\be
\label{DCDC}
\begin{aligned}
{\cal L}_{3-form} = & \left( \partial_m C^m \right)^2 + \left( 2i A A^\Psi \partial_m C^m + c.c. \right) \\ &- 2 \partial_m \left( C^m \partial_n C^n \right) - i\partial_m (C^m(2 A A^\Psi - 2 \overline{A} \overline{A}^\Psi)) \, , 
\end{aligned}
\ee 
where $A=X|$, $A^\Psi=\Psi|$ and the total derivative terms are the relevant part of the boundary term \eqref{BD}.
Here we refer to $C_m$ as the gauge three-form because, even though $C_m$ is a 1-form, 
it is in fact the Hodge-dual of the gauge three-form $C_3$ (see \eqref{Cm}).
We also point out that the boundary term \eqref{BD} actually contains many more total derivatives in addition to those in the second line of \eqref{DCDC}. However, these contributions to $\mathcal{L}_{\text{bd}}$ can be ignored since they do not involve $C_m=(\star C_3)_m$ and are therefore not necessary for the correct definition of the variational problem with respect to the gauge three-form as pointed out for example in \cite{Lanza:2019xxg}.

Thanks to the boundary term in \eqref{DCDC}, we can vary the action $S_{X-\Psi}=\int d^4 x \, \mathcal{L}_{X-\Psi}$ with respect to $C_m$ subject to the boundary condition \cite{Groh:2013} 
\be
\label{bc}
\delta (\star d C_3)\Big{|}_{\text{Boundary}}= \delta (\partial_m  C^m) \Big{|}_{\text{Boundary}} = 0 \, , 
\ee
which is invariant under the bosonic gauge transformation $C_3 \to C_3 + dB_2$ contained in the superfield transformation \eqref{UL}, to get the following equation of motion 
\be
\label{fluxGLOB}
\partial_m (\partial_n C^n + i A A^\Psi - i \overline{A} \overline{A}^\Psi) = 0 \ \ \to \ \ \partial_n C^n  + i A A^\Psi - i \overline{A} \overline{A}^\Psi = {\rm n} \, , 
\ee 
where n is a real constant. 
Plugging the above equation into the gauge three-form sector of \eqref{nil-3form} we obtain
\be
{\cal L}_{3-form} = - ({\rm n} - i A A^\Psi + i \overline{A} \overline{A}^\Psi)^2 \, . 
\ee
Furthermore, the terms of \eqref{nil-3form} that include the real scalar $\mathcal{F} $ are 
\be
{\cal L}_{\cal F}={\cal F}^2+2 \mathcal {F} (A A^\Psi + \overline{A} \overline{A}^\Psi ) \, ,
\ee
from which we can easily deduce that the Euler-Lagrange equation for $\mathcal{F}$ reads
\be
\label{ReF}
\mathcal {F}=-(A A^\Psi + \overline{A} \overline{A}^\Psi) \, . 
\ee
We may also use \eqref{F3form} to combine equations \eqref{fluxGLOB} and \eqref{ReF} into
\be
\label{F}
F=i {\rm n} - 2 \overline{A} \overline{A}^\Psi\,. 
\ee 
Using the standard supersymmetry formulas found in \cite{Wess:1992cp} we find the full component form of \eqref{nil-3form} after substituting \eqref{fluxGLOB} and \eqref{ReF} into it, which reads
\be
\label{Lcomp}
\begin{aligned}
 \mathcal{L}_{X-\Psi} = & \overline{A} \partial^2 A + i \partial_m \overline{G} \overline{\s}^m G + (A^2 F^{\Psi} - 2 A G \psi - A^\Psi GG +c.c.) 
 \\
 &-n^2+2in(A A^\Psi - \overline{A} \overline{A}^\Psi)-4|A|^2|A^\Psi|^2 \,,
\end{aligned}
\ee
where $\psi_{\alpha}=\frac{1}{\sqrt{2}} D_{\alpha} \Psi |$ and $F^{\Psi}=-\frac{1}{4} D^2 \Psi |$. Our next task is to integrate out the components $A^\Psi$, $\psi_\alpha$ and $F^\Psi$ of the Lagrange multiplier chiral superfield $\Psi$, which leads to
\be
\label{Psi eom}
\begin{aligned}
\delta A^\Psi &: \ \ GG=2A(in - 2 \overline{A} \overline{A}^\Psi) \overset{\eqref{F}}{=} 2AF \, , 
\\
\delta \psi_\alpha &: \ \ A G^\alpha=0 \,, 
\\
\delta F^\Psi &: \ \ A^2=0 \,.
\end{aligned}
\ee
Equations \eqref{Psi eom} are equivalent to the nilpotency constraint $X^2=0$ (see \eqref{X2zero}). 
Plugging them into \eqref{Lcomp} and making use of \eqref{F}, we get the Lagrangian
\be
\label{LX}
\mathcal{L}_X =  \overline{A} \partial^2 A + i \partial_m \overline{G} \overline{\s}^m G + F \overline{F} + (i {\rm n} F + c.c.) \,, 
\ee
where $X=A(y)+\sqrt{2}\theta G(y)+\theta^2 F(y)$ with $y^m \equiv x^m + i \theta \s^m \overline{\theta}$ \cite{Wess:1992cp} satisfies 
\be
\label{X20}
X^2=0 \,.
\ee
We identify \eqref{LX} as the component form of the superspace Lagrangian \eqref{LXPsi} after the chiral Lagrange multiplier $\Psi$ has been integrated out to impose the condition $X^2=0$ but with the real $f$ replaced by the imaginary constant $i \rm n$. We can thus argue that the solution to the constraint \eqref{X20} depends here on whether n takes a non-vanishing value or not. We have
\be
\label{solX20}
{\rm I)} \ {\rm n} \ne 0 \ \to \  A = \frac{G^2}{2F}= \frac{G^2}{2 (\F + i \partial_m C^m)} \, 
\quad 
{\text{or}}\,
\quad 
{\rm II)} \ {\rm n} = 0 \ \to \  X = 0 \, ,
\ee 
which implies that 
\be 
\label{nis0}
{\cal L}_{X,{\rm n} = 0} = 0 \, , 
\ee
while
\be
\label{nne01}
{\cal L}_{X,{\rm n} \ne 0} =\frac{\overline{G}^2}{2\overline{F}} \partial^2 \left( \frac{G^2}{2F} \right) + i \partial_m \overline{G} \overline{\s}^m G + F \overline{F} + (i {\rm n} F + c.c.) \,. 
\ee
In the case where n$\ne0$, the equations of motion of the auxiliary fields $F$, $\overline{F}$ are
\be
\label{Feom}
F-\frac{\overline{G}^2}{2\overline{F}^2} \partial^2 \left( \frac{G^2}{2F} \right)-i {\rm n}=0 \ , \quad 
\overline{F}-\frac{G^2}{2F^2} \partial^2 \left( \frac{\overline{G}^2}{2\overline{F}}\right) +  i {\rm n}=0 \,.
\ee
The solution of the above equations is given by 
\be
\label{Fsol}
\begin{aligned}
F&=i {\rm n}\left( 1 + \frac{1}{4 {\rm n}^4 } \overline{G}^2 \partial^2 G^2 - \frac{3}{16 n^8} G^2 \overline{G}^2 \partial^2 G^2  \partial^2 \overline{G}^2 \right) \,, 
\\
\overline{F}&= - i {\rm n}\left( 1 + \frac{1}{4 {\rm n}^4 } G^2 \partial^2 \overline{G}^2 - \frac{3}{16 n^8} G^2 \overline{G}^2 \partial^2 G^2  \partial^2 \overline{G}^2 \right) \,,
\end{aligned}
\ee
which, substituted into \eqref{nne01}, gives 
\be
\label{nne0}
{\cal L}_{X,{\rm n} \ne 0} = -{\rm n}^2 + i \partial_m \overline G \overline \sigma^m G 
+ \frac{1}{4 {\rm n}^2} \overline G^2 \partial^2 G^2 
-\frac{1}{16 {\rm n}^6} G^2 \overline G^2 \partial^2 G^2 \partial^2 \overline G^2 \,.
\ee
From the solution \eqref{Fsol} it is clear that for ${\rm n}\ne0$ and a Lorentz invariant vacuum we have that 
\be
\label{Fvev}
\langle F \rangle = i {\rm n} \quad \overset{\eqref{F3form}}{\Longleftrightarrow} \quad 
\langle \mathcal{F} \rangle =0 \ {\text{and}} \ \langle \partial_m C^m\rangle={\rm n} \,,
\ee
so n represents a flux background. Equations \eqref{Fvev} hold trivially for ${\rm n} = 0$ as well. 

If we find a way to have transitions between 
${\rm n}\ne0$ and ${\rm n}=0$ then this means we will have a transition between \eqref{nne0} and \eqref{nis0}. 
In this way the goldstino fermion, which is propagating in \eqref{nne0}, while supersymmetry 
is spontaneously broken and intrinsically non-linearly realized, 
will evaporate in \eqref{nis0} and supersymmetry will be restored (at least in this sector). 
Now we will turn to the coupling to super-membranes that allow such transitions.

The membrane action we consider has the form 
\be
\label{MEMG}
S_{membrane} = - \frac{1}{4 \pi} \int_{\mathcal{M}_3} d^3 \xi \sqrt{-\text{det}\hspace{0.05cm} h_{ij}} |\mu X + c| 
-\frac{\mu}{4 \pi} \int_{\mathcal{M}_3} {\cal C}_3 
- \left( \frac{\overline c}{4 \pi} \int_{\mathcal{M}_3} {\cal C}_3^0 + c.c. \right) \,. 
\ee
Equation \eqref{MEMG} gives the most general action describing the coupling of the single three-form chiral superfield $X$ to a membrane in flat 4D N=1 superspace with supercoordinates $z^M=(x^m,\theta^\alpha,\overline{\theta}_{\dot{\alpha}})$ \cite{Bandos:2019qok}. $\mathcal{M}_3$ denotes the membrane world-volume, which is parametrized by coordinates $\xi^i, i=0,1,2,$ and whose embedding into the background flat superspace is described by equations of the form $z^M=z^M(\xi)$, and the superfield $X(x,\theta,\overline{\theta})$ is evaluated at $z^M=z^M(\xi)$.
Furthermore,
\be
h_{ij}(\xi)\equiv\eta_{ab}E^a_i(\xi)E^b_j(\xi)\,, \ \ \text{where} \ \ E^a_i(\xi) \equiv \p_iz^M(\xi) E_M^a(z(\xi))\,, 
\ee
is the induced metric on the membrane world-volume and using the explicit expressions for the supervielbein of flat 4D N=1 superspace given in \cite{Wess:1992cp} one can derive the pullback of the vector supervielbein $E^a=dz^ME_M^a$ on $\mathcal{M}_3$, which reads
\be
E^a(\xi)\equiv dz^M(\xi)E_M^a(z(\xi))=d\xi^iE^a_i(\xi)=dx^a(\xi) - id\theta(\xi)\sigma^a\overline{\theta}(\xi) + i\theta(\xi)\sigma^ad\overline{\theta}(\xi)\,.
\ee
The constants $\mu$ and $c$ are a real and a complex charge respectively that characterise the coupling of the membrane to a real super three-form $\mathcal{C}_3$ and a complex super three-form $\mathcal{C}_3^0$, where
 $\mathcal{C}_3$ is defined in terms of the real prepotential $U$ by \cite{Gates:1980ay}
\begin{align}
\mathcal{C}_3 \equiv  &  iE^a\wedge d\theta^\alpha \wedge d{\overline{\theta}}^{\dot{\alpha}} \s_{a\hspace{0.05cm}\alpha\dot{\alpha}} U - \frac{1}{2} E^b \wedge E^a \wedge d\theta^\alpha {\s_{ab \hspace{0.05cm} \alpha}}^\beta D_\beta U  \nonumber \\
 &-\frac{1}{2} E^b \wedge E^a \wedge d\overline{\theta}^{\dot{\alpha}} 
 {{\overline{\s}_{ab}}^{\dot{\beta}}}_{\dot{\alpha}} \overline{D}_{\dot{\beta}}U - \frac{1}{48}E^c \wedge E^b \wedge E^a \epsilon_{abcd} \overline{\s}^{d \hspace{0.05cm} \dot{\alpha} \alpha} [D_\alpha,\overline{D}_{\dot{\alpha}}] U\,,
\end{align}
while $\mathcal{C}_3^0$ is given by
\be
C_3^0=iE^a \wedge d\theta^\alpha \wedge d{\overline{\theta}}^{\dot{\alpha}}\s_{a\hspace{0.05cm}\alpha\dot{\alpha}}\theta^2 - E^b \wedge E^a \wedge d\theta^\alpha {\s_{ab \hspace{0.05cm} \alpha}}^\beta \theta_\beta\,.
\ee

By construction, the action \eqref{MEMG} is invariant under reparametrizations of the membrane world-volume $\xi^i \rightarrow \xi'^i(\xi)$ as well as under the $\kappa$-symmetry transformations 
\be
\delta_\kappa \theta^\alpha (\xi)=\kappa^\alpha(\xi)\,, \ \ \delta_\kappa \overline{\theta}^{\dot{\alpha}}(\xi)= \overline{\kappa}^{\dot{\alpha}} (\xi)\equiv(\kappa^\alpha(\xi))^*\,, \ \
\delta_\kappa x^m(\xi)= i\kappa(\xi)\s^m\overline{\theta}(\xi)-i\theta(\xi)\s^m\overline{\kappa}(\xi)\,,
\ee
which satisfy
\be
\label{i_k E}
i_\kappa E^a  \equiv \delta_\kappa z^M E_M^a(z)=0\,, \ \ 
i_\kappa E^\alpha  \equiv \delta_\kappa z^M E_M^\alpha(z)=\kappa^\alpha\,, \ \  
i_\kappa \overline{E}^{\dot{\alpha}} \equiv  \delta_\kappa z^M E_M^{\dot{\alpha}}(z)=\overline{\kappa}^{\dot{\alpha}}\,.
\ee
The local fermionic parameter $\kappa^{\alpha}(\xi)$ obeys the projection condition 
\be
\kappa_{\alpha}=-i\frac{\mu X + c}{|\mu X + c|}\Gamma_{\alpha \dot{\alpha}}\overline{\kappa}^{\dot{\alpha}} \,,
\ee
where 
\be
\label{Gamma}
\Gamma_{\alpha \dot{\alpha}} \equiv \frac{\text{i} \epsilon^{ijk}}{3!\sqrt{-\text{det}\hspace{0.05cm}h}} \epsilon_{abcd} E^b_i E^c_j E^d_k \s^a_{\alpha \dot{\alpha}}\,, 
\ee 
and defining $\overline{\Gamma}^{\dot{\alpha} {\alpha}} \equiv \epsilon^{\alpha \beta} \epsilon^{\dot{\alpha} \dot{\beta}} \Gamma_{\beta \dot{\beta}}$ we have that 
\be 
\Gamma_{\alpha \dot{\alpha}} \overline{\Gamma}^{\dot{\alpha}{\beta}}=\delta_{\alpha}^{\beta}\,.
\ee

Let us now consider the total action (bulk + membrane) of the form 
\be
S = S_{X-\Psi} + S_{membrane} \, , 
\ee
which leads to a bosonic sector 
\be
\label{bosMEM+X}
\begin{aligned}
S_{\text{bos}} = & \int d^4 x \, \left\{ \mathcal{F}^2 + \left(  \partial_m C^m \right)^2 - 2 \partial_m \left( C^m \partial_n C^n \right) \right\} \\
& - \frac{|c|}{4 \pi}  \int_{\mathcal{M}_3} d^3 \xi \, \sqrt{- \text{det} \, (\eta_{ab} \partial_i x^a(\xi) \partial_j x^b(\xi)  )}  
- \frac{\mu}{4 \pi}  \int_{\mathcal{M}_3} C_3 \, , 
\end{aligned}
\ee
where we have already set $A=0$ because according to \eqref{solX20}, $A$ will either vanish identically or involve the fermion bilinear $G^2$, depending on the value of the flux ${\rm n} = \langle \partial_m C^m \rangle $,
and the relation between the $C_3$ and $C_m$ is 
\be
\label{C3C1}
C_3 = \frac{1}{3!} dx^c \wedge dx^b \wedge dx^a C_{abc}=\frac{1}{3!} dx^c \wedge dx^b \wedge dx^a \epsilon_{abcd} C^d \, . 
\ee
In \eqref{bosMEM+X} we have the pullback of $C_3$ on the membrane world-volume 
\be
\int_{\mathcal{M}_3} C_3 = \frac{1}{3!} \int_{\mathcal{M}_3} d\xi^k\wedge d\xi^j \wedge d\xi^i \, \partial_k x^c (\xi) \partial_j x^b (\xi) \partial_i x^a (\xi)\, C_{abc} (x(\xi)) \, . 
\ee 
We can now evaluate the flux jump following for example \cite{Bandos:2019wgy}. 
Varying the action \eqref{bosMEM+X} with respect to $C_m$ subject to the boundary condition \eqref{bc} we obtain the following equation of motion
\be
\label{Ceom}
\partial_m \partial_n C^n = - \frac{\mu}{48 \pi} \int_{\mathcal{M}_3} d^3 \xi \, \epsilon^{kji} \partial_k x^q(\xi) \partial_j x^p(\xi) \partial_i x^n(\xi) \epsilon_{npqm} \delta^{(4)} (x-x(\xi)) \, ,
\ee
where we have used \eqref{C3C1} and $d\xi^k\wedge d\xi^j \wedge d\xi^i \equiv \epsilon^{kji} d^3 \xi$. Equation \eqref{Ceom} implies that away from the membrane, where $x^m \ne x^m(\xi)$, $\partial_m C^m = \star dC_3$ 
is constant since the delta function on its right hand side vanishes.
If the membrane world-volume is closed, it will divide the background flat superspace into an inside and an outside region and from \eqref{Ceom} it follows that the vacuum expectation value of $\partial_m C^m$ will be different in these two regions. 
In order to compute the difference between ${\rm n}_o \equiv \langle \partial_m C^m \rangle |_{\text{outside}}$ and ${\rm n}_i \equiv \langle \partial_m C^m \rangle |_{\text{inside}}$, 
we consider a small neighborhood of a point on the membrane where we impose the static gauge 
\be
\label{static}
x^i(\xi)=\xi^i, \ \ i=0,1,2, 
\ee
which gauge fixes the world-volume reparametrization invariance of the action \eqref{MEMG}. Then, in this neighborhood, the membrane world-volume is described by $x^3(\xi)=0$ and the $m=3$ component of \eqref{Ceom} reads
\be
\label{3comp}
\partial_3 \partial_n C^n = - \frac{\mu}{8 \pi} \delta(x^3) \int d^3 \xi \delta(x^0-\xi^0) \delta(x^1-\xi^1) \delta(x^2-\xi^2) 
= - \frac{\mu}{8 \pi} \delta(x^3) \, . 
\ee
Integrating the above equation with respect to $x^3$ over the interval $[-\epsilon,\epsilon]$, where $\epsilon$ is an infinitesimal positive real number, we obtain
\be
\label{Dn}
\Delta {\rm n} \equiv {\rm n}_o - {\rm n}_i=- \frac{\mu}{8 \pi} \,. 
\ee
Therefore, if ${\rm n}_o=f$, where $f$ is a non-zero real constant and  $\mu=-8\pi f$, then ${\rm n}_i=0$, which means that
the value of ${\rm n} =\langle \partial_m C^m \rangle $ goes from non-vanishing to vanishing and so there is a 
transition between $A = G^2 /2F$ and $X\equiv0$. 
Then as far as \eqref{MEMG} is concerned, 
after the restoration of supersymmetry the goldstino evaporates ($X \equiv 0$) 
and one has a free membrane on a Minkowski background.

\section{Decay of pure de Sitter supergravity}

In this section we discuss the decay channel that is related to the goldstino evaporation in a supergravity setup. 
On a conceptual level, 
the evaporation of the full goldstino superfield happens essentially in the same way as we have described in the 
global supersymmetry section so we do not need to go through it again. 
The only difference is a few technical difficulties that appear due to the more involved structure of 
the membrane and 3-form couplings in supergravity 
(some recent developments on the topic can be found for example in \cite{Bielleman:2015ina,Bandos:2018gjp,Lanza:2019xxg,Bandos:2019lps}). 
We will still verify that the transition happens in a discrete jump and then we will study the life-time of the de Sitter phase.

\subsection{Supergravity and 3-forms} 
In the old-minimal 4D N=1 supergravity, a general chiral superfield $\Phi$  
satisfies the constraint $\overline{\mathcal{D}}_{\dot{\alpha}} \Phi = 0$ and its component fields can be defined by projection as 
\be
\phi=\Phi|, \ \ \chi_\alpha=\frac{1}{\sqrt{2}} \mathcal{D}_{\alpha} \Phi|, \ \ F = - \frac{1}{4} \mathcal{D}^2 \Phi | \, ,
\ee
where $\phi$ is a complex scalar, $\chi_\alpha$ is a Weyl fermion and $F$ is a complex auxiliary scalar. The chiral superfield $\Phi$ can also be written as a power series in the new fermionic variables $\Theta^\alpha$ \cite{Wess:1992cp} as 
$\Phi=\phi(x) + \sqrt{2} \Theta \chi (x) + \Theta^2 F(x)$. 
The most general 2-derivative locally supersymmetric Lagrangian describing a system of chiral superfields $ \Phi^i $ coupled to supergravity is (we set the gravitational coupling constant $\kappa$ equal to one) 
\be
\label{LSG chiral}
\mathcal{L}=-3 \int d^4 \theta \, E e^{-\frac{K(\Phi^i,\overline{\Phi}^{\overline{i}})}{3}} + \left( \int d^2 \Theta \, 2 \mathcal{E} W(\Phi^i) + c.c. \right) \, ,
\ee
where $E$ is the superdeterminant of the supervielbein $E_M^A$, $2\mathcal{E}$ is the chiral density defined in \cite{Wess:1992cp}, $K(\Phi^i,\overline{\Phi}^{\overline{i}})$ is a real function of the superfields $ \Phi^i $ and $\overline{\Phi}^{\overline{i}} \equiv(\Phi^i)^*$ called the K\"ahler potential and $W(\Phi^i)$ is a holomorphic function of the $ \Phi^i $ known as the superpotential.

 A single three-form chiral superfield $\mathcal {X}$ can be  expressed in terms of 
a real superfield prepotential $U$  \cite{Binetruy:1996,Ovrut:1997ur,Bandos:2012gz} as 
\be
\label{sugra-X}
{\cal X} = - \frac14 \left(\overline{\cal D}^2 -8 {\cal R}\right) U \, ,
\ee
where a bosonic three-form $C_3$ resides in the vector component of $U$ according to
\be
\label{bosC3}
 - \frac18 \overline \sigma^{ \dot \alpha \alpha}_a [{\cal D}_\alpha , \overline{\cal D}_{\dot \alpha} ] U | = (\star C_3)_a \equiv 
  \frac{1}{3!} \epsilon_{abcd} C^{bcd} \equiv C_a\, , 
\ee
 and $\mathcal{R}$ is the Ricci chiral superfield, which has lowest 
component ${\cal R}| = - M /6$ with $M$ being the complex auxiliary scalar field of the supergravity multiplet. 
We note that \eqref{sugra-X} is invariant under the superfield gauge transformation 
\be
U \to U+L \, ,
\ee
where $L$ is a real linear superfield satisfying the constraint $(\overline{\mathcal{D}}^2 - 8\mathcal{R})L=0$. By an appropriate choice of $L$ the real prepotential $U$ can be put into a WZ gauge where \cite{Binetruy:1996,Ovrut:1997ur}
\be
U| = \mathcal{D}_{\alpha}U| = \overline{\mathcal{D}}_{\dot{\alpha}}U| = 0 \, .
\ee
Without loss of generality, we can compute the auxiliary component field of $\mathcal{X}$, $F=-\frac{1}{4}\mathcal{D}^2 \mathcal{X}|$, in this gauge to find that the corresponding bosonic sector reads 
\be
\label{F-bosonic}
F|_{\text{bos}} = {\cal F} + i D_a C^a + \frac12 \overline M A - \frac12 M \overline A \, , 
\ee
where $\mathcal{F}\equiv \operatorname{Re} F$ is a real auxiliary scalar, $A=\mathcal{X}|$ and  $D_a C^a = e_a^m D_m C^a = e_a^m (\partial_m C^a + C^b {\omega_{mb}}^a(e))$ where ${\omega_{mb}}^a(e)$ is the spin connection defined as in \cite{Wess:1992cp} without the terms involving the gravitino $ {\psi_m}^\alpha $.

Following \cite{Bandos:2018gjp} we consider the following action describing the coupling of the single three-form chiral superfield $\mathcal{X}$ to a membrane in a curved 4D N=1 superspace with supercoordinates $z^M=(x^m,\theta^\mu,\overline{\theta}_{\dot{\mu}} )$ 
\be
\label{MEMG-SG}
S_{membrane} = - \frac{1}{4 \pi} \int_{\mathcal{M}_3} d^3 \xi \sqrt{-\text{det}\hspace{0.05cm} h_{ij}} |\mu {\cal X}| 
-\frac{\mu}{4 \pi} \int_{\mathcal{M}_3} {\cal C}_3  \,, 
\ee 
where $\mathcal{M}_3$ is the membrane world-volume with coordinates $\xi^i, i=0,1,2,$ whose embedding into the background superspace is described by parametric equations of the form $z^M=z^M(\xi)$, and the bulk superfield $ \mathcal{X}(x,\theta,\overline{\theta})$ is evaluated on $\mathcal{M}_3$. Furthermore, $h_{ij}(\xi)\equiv\eta_{ab}E^a_i(\xi)E^b_j(\xi) $ with $E^a_i(\xi) \equiv \p_iz^M(\xi) E_M^a(z(\xi)) $ is the induced metric on the membrane world-volume and $ \mu $ is a real constant charge characterising the membrane coupling to a real super three-form
$\mathcal{C}_3 $ defined in terms of $U$ by 
\begin{align}
\mathcal{C}_3 \equiv  &  iE^a\wedge E^\alpha \wedge \overline{E}^{\dot{\alpha}} \s_{a\hspace{0.05cm}\alpha\dot{\alpha}} U - \frac{1}{2} E^b \wedge E^a \wedge E^\alpha {\s_{ab \hspace{0.05cm} \alpha}}^\beta \mathcal{D}_\beta U  \nonumber \\
 &-\frac{1}{2} E^b \wedge E^a \wedge \overline{E}^{\dot{\alpha}} {{\overline{\s}_{ab}}^{\dot{\beta}}}_{\dot{\alpha}} \overline{\mathcal{D}}_{\dot{\beta}}U - \frac{1}{48}E^c \wedge E^b \wedge E^a \epsilon_{abcd} (\overline{\s}^{d \hspace{0.05cm} \dot{\alpha} \alpha} [\mathcal{D}_\alpha,\overline{\mathcal{D}}_{\dot{\alpha}}] U + 8 G^d U) \,. 
\end{align}
The action \eqref{MEMG-SG} is invariant under world-volume diffeomorphisms $ \xi^i \to \xi'^i(\xi)$ 
as well as under the $\kappa$-symmetry transformations 
\be
\delta_{\kappa} z^M (\xi) = \kappa^\alpha(\xi) E_\alpha^M (z(\xi)) + \overline{\kappa}_{\dot{\alpha}} (\xi) E^{\dot{\alpha}M}(z(\xi)) \, ,
\ee
which satisfy equations \eqref{i_k E} and the local fermionic parameter $\kappa^\alpha (\xi)$ satisfies the projection condition
\be
\kappa_{\alpha}=-i\frac{\mu \mathcal{X} }{|\mu \mathcal{X} |}\Gamma_{\alpha \dot{\alpha}}\overline{\kappa}^{\dot{\alpha}} \,, 
\ee
where $ \Gamma_{\alpha \dot{\alpha}}$ is defined in \eqref{Gamma}.

\subsection{Discrete jumps in de Sitter supergravity with 3-forms}

Now we can study the decay channel for the so-called pure de Sitter supergravity \cite{Bergshoeff:2015tra}. 
For a standard nilpotent chiral superfield the K\"ahler potential, 
the superpotential and the nilpotency constraint read 
\be
K = |X|^2 \ , \quad W = f X + m_{3/2} \ , \quad X^2=0 \,, 
\ee
where we can have $m_{3/2}$ and $f$ real constants, 
which lead to a complete bosonic sector of the form 
\be 
\label{DSsugra}
e^{-1} {\cal L}_{dS-SUGRA} = - \frac12 R - f^2 + 3 m_{3/2}^2 \, . 
\ee
In our setup the constant $f$ is generated dynamically as a flux and we want to study transitions between 
$f \ne 0$ and $f=0$. 
To do this we have to recast the pure de Sitter supergravity in terms of a nilpotent three-form chiral superfield, 
which will be coupled to a membrane.

At this point however we face a small technical difficulty.\footnote{
In contrast to \cite{Farakos:2016hly} here we have fully kappa-symmetric super-membranes 
and so this new complication is linked to supersymmetry.} 
From \eqref{MEMG-SG} we see that the lowest component of the chiral superfield ${\cal X}$ 
combines with $\mu$ to give rise to the tension. 
As a result we cannot set ${\cal X}^2=0$ because this will give rise to a vanishing effective membrane tension. 
At the same time we cannot employ another independent three-form chiral superfield, say $Y$, and then set $Y=$const.,   
such that we can build an independent background membrane as we did in global supersymmetry, 
because such a procedure will instead restrict the auxiliary fields of the supergravity multiplet 
and lead to the so-called three-form supergravity \cite{Ovrut:1997ur}, 
forcing the three-form that appears in the WZ term to become the supergravity three-form. 
The solution to this difficulty lies in between the two: 
we have to impose a condition of the form ${\cal X} =$const. via a nilpotency condition such 
that no other components of ${\cal X}$ are restricted. 
To this end we will make use of the shifted nilpotency condition 
\be 
\label{ShiftedX}
\left( {\cal X} + \frac{4 \pi T}{\mu} \right)^2=0 \, , 
\ee 
which is solved by
\be
{\cal X} \Big{|}_{SUSY \ broken} = - \frac{4  \pi T}{\mu} + \frac{G^2}{2 F} + \sqrt 2 \Theta G + \Theta^2 F \, , 
\ee
thus giving $\langle {\cal X}| \rangle = - 4  \pi T / \mu$. 
Note that because this is a three-form multiplet the $F$ component contains the real scalar ${\cal F}$ 
in its real part and the four-form field strength in its imaginary part, 
but also the auxiliary field $M$ appears (see e.g. \cite{Bandos:2018gjp}) together with fermionic terms. 
To be precise, for the bosonic terms we have 
\be
\label{F-bosonic-nilpotent}
F|_{bos} = {\cal F} + i D^a C_a + \frac{4 i  \pi T}{\mu} {\rm Im} M  \,. 
\ee
Here both $T$ and $\mu$ are real and they have mass dimensions $[T] = 3$ and $[\mu] = 2$, 
while we remain always in Planck units. 
When 
 supersymmetry is restored, the shifted nilpotency condition has only the trivial solution 
\be
{\cal X} \Big{|}_{SUSY \ restored} \equiv - \frac{4  \pi T}{\mu} \, , 
\ee
which just brings us to a SUSY AdS 3-form supergravity.

The full supergravity+membrane action we will study is 
\be 
\label{TOT-ACTION}
S = S_{SUGRA} + S_{boundary} + S_{membrane} \, , 
\ee 
with the latter given by \eqref{MEMG-SG} 
whereas for the supergravity sector we have a K\"ahler potential and a superpotential given 
by\footnote{The self-consistency of imposing nilpotency constraints with Lagrange multipliers also in supergravity 
has been carefully analyzed in \cite{Ferrara:2016een}.} 
\be
K = \Big{|} {\cal X} + \frac{4 \pi T}{\mu} \Big{|}^2 \ , 
\quad W = m_{3/2} + \Psi \left( {\cal X} + \frac{4 \pi T}{\mu} \right)^2 \, . 
\ee
$S_{boundary}$ contains the terms that are essential for the consistent variation of \eqref{TOT-ACTION} with respect to the three-form 
and we will always assume that they are present from now on without writing them down explicitly. 
We will now focus on the bosonic sector and verify the above setup can give the Lagrangian \eqref{DSsugra} 
and that the membrane enables transitions from $f \ne 0$ to $f=0$. 
Later we give the full superspace equivalence, which also provides the 
exact identification of the supersymmetry breaking scale. 
The bulk bosonic sector (after we integrate out the $b_a$ auxiliary field of the supergravity multiplet) reads 
\be
\begin{aligned}
e^{-1} {\cal L}_{bos-bulk} =& \frac{1}{6} \left( \left| A + \frac{4 \pi T}{\mu} \right|^2  - 3  \right) R
+ |F|^2 
+ 2 F A^\Psi \left(A + \frac{4 \pi T}{\mu} \right) 
+ 2 \overline F \overline A^\Psi \left(\overline A + \frac{4 \pi T}{\mu} \right) 
\\
& 
- \frac13 {\rm Re} M^2 
- \frac13 {\rm Im} M^2 
+ F^\Psi \left( A + \frac{4 \pi T}{\mu} \right)^2 
+ \overline F^\Psi \left( \overline A + \frac{4 \pi T}{\mu} \right)^2 
\\
&- \overline M \left( m_{3/2} + A^\Psi \left( A + \frac{4 \pi T}{\mu} \right)^2 \right) 
- M \left( m_{3/2} + \overline A^\Psi \left( \overline A + \frac{4 \pi T}{\mu} \right)^2 \right) 
\\
& + \frac{1}{9} \left| A + \frac{4 \pi T}{\mu} \right|^2 |M|^2 -\frac{1}{3} M \left(\overline A + \frac{4 \pi T}{\mu} \right) F -\frac{1}{3} \overline{M} \left( A + \frac{4 \pi T}{\mu} \right) \overline{F} -\partial_m A \partial^m \overline{A}
\\
&  - \frac{1}{4} \left( \left| A + \frac{4 \pi T}{\mu} \right|^2 - 3 \right)^{-1} \left[ \left( \overline A + \frac{4 \pi T}{\mu} \right) \partial_m A - \left(  A + \frac{4 \pi T}{\mu} \right) \partial_m \overline{A} \right]^2 \, , 
\end{aligned}
\ee
where $F$ is given by \eqref{F-bosonic}. 
Now let us integrate out the easy auxiliary fields and the Lagrange multiplier sector, 
keeping in mind that these fields do not enter the bosonic membrane sector. 
First we integrate out $F^\Psi$ which gives 
\be
A = - \frac{4  \pi T}{\mu} + {\text{fermions}} \, , 
\ee
and turns $F$ into \eqref{F-bosonic-nilpotent}. 
Then the equation of motion for $A^\Psi$ just becomes a trivial consistency condition which is automatically satisfied, 
and now we can easily integrate out ${\rm Re} M$ to find the usual $3 m_{3/2}^2$ contribution, 
keeping in mind that $m_{3/2}$ is a real constant. 
Similarly the Euler-Lagrange equation for ${\cal F}$ just gives ${\cal F}=0$. 
The bulk bosonic sector thus reduces to 
\be
e^{-1} {\cal L}_{ bos-bulk} = - \frac12 R 
- \frac13 {\rm Im} M^2 
+ \left( D^a C_a + \frac{4  \pi T}{\mu} {\rm Im} M \right)^2 
+ 3 m_{3/2}^2 \,. 
\ee
Finally, we integrate out ${\rm Im} M$ and we have 
\be
{\rm Im} M = \frac{4 \pi T}{\mu} \left( \frac13 - \left( \frac{4 \pi T}{\mu} \right)^2 \right)^{-1} D^a C_a \,, 
\ee
which leads to the following form of the bulk Lagrangian 
\be
\label{final-bulk}
e^{-1} {\cal L}_{ bos-bulk} = - \frac12 R 
+ \frac{(D^a C_a)^2 }{1 - 3 \left( \frac{4 \pi T}{\mu} \right)^2} 
+ 3 m_{3/2}^2 
- 2 D^b \left( C_b \, \frac{D^a C_a}{1 - 3 \left( \frac{4 \pi T}{\mu} \right)^2} \right) \,, 
\ee  
where we also included the appropriate boundary term. 
The bosonic sector of the membrane action reads
\be
\label{super-membrane-SG}
S_{bos-membrane} = - |T| \int_{\mathcal{M}_3} d^3 \xi \sqrt{- \text{det}\hspace{0.05cm} \gamma_{ij}} 
- \frac{\mu}{4 \pi}   \int_{\mathcal{M}_3} C_3 \, , 
\ee
where $\gamma_{ij}(\xi) \equiv g_{mn} (x(\xi)) \partial_i x^m(\xi) \partial_j x^n(\xi)$ is the pull-back of the bulk metric on the membrane  world-volume and the bosonic three-form $C_3$ is defined in \eqref{bosC3}. 
Note that from \eqref{super-membrane-SG} the tension $T_3$ of the membrane is 
\be
T_3 = |T| >0 \, , 
\ee
and because of the pre-factor appearing in \eqref{final-bulk} we can define a three-form with canonically 
normalized kinetic term 
\be
\tilde C_a = \left(1 - 3 \left( \frac{4 \pi T}{\mu} \right)^2\right)^{-1/2} C_a \, . 
\ee
This gives a WZ term of the form $Q \int_{\mathcal{M}_3} \tilde C_3$, 
where
\be
\label{Q-def}
Q = - \frac{\mu}{4 \pi}  \sqrt{1 - 3 \left( \frac{4 \pi T}{\mu} \right)^2}   \, , 
\ee
is the effective membrane charge under the canonical 3-form, 
which also enters the jump induced on the flux by the membrane.
From now on we also set 
\be
Q>0 \ , \quad \mu <0 \,. 
\ee
From the definition of the charge \eqref{Q-def} we can also express $\mu$ completely in terms of the former and the tension 
\be
\label{muQT}
\mu = - 4 \pi \sqrt{Q^2 + 3 T^2} \, . 
\ee

Once we integrate out $\tilde C_3$ we find the usual contribution to the cosmological constant. 
We have to note that the contribution from the boundary terms is crucial as it guarantees the correct sign of the final result. 
The full bulk bosonic sector then reads 
\be
\label{boson-3forms}
e^{-1} {\cal L}_{ bos-bulk} = - \frac12 R - {\rm n}^2 + 3 m_{3/2}^2  \, , 
\ee
where ${\rm n} = \langle D^a \tilde C_a \rangle >0$ is the flux in terms of the canonically normalized three-form. 
When we have membrane nucleation the value of the cosmological constant changes as 
\be
\Lambda_o = {\rm n}_o^2 - 3 \, m_{3/2}^2 \ \ \to \ \ \Lambda_i = \left({\rm n}_o - \Delta n \right)^2 - 3 \, m_{3/2}^2 \, , 
\ee
and the jump is given by 
\be
\Delta {\rm n} = \frac{Q}{2} \, . 
\ee

\subsection{The supersymmetry breaking scale and decay rate}

At this point we need to discuss the supersymmetry breaking scale, 
which defines the cut-off, 
and thus relates the membrane charge to the tension. 
We do not need to discuss the membrane sector because it only induces jumps in the flux value. 
We thus focus on the bulk sector and work with the superspace Lagrangian 
\be
\label{SG-EQUIV}
\begin{aligned}
{\cal L} = & \int d^4 \theta E \left( - 3 \, e^{- \frac13 \Big{|} {\cal X} + \frac{4 \pi T}{\mu} \Big{|}^2} 
- U (Z + \overline Z) \right) 
\\
& + \left[ 
\int d^2 \Theta \, 2 {\cal E} \left( m_{3/2} + Z {\cal X} + \Psi \left( {\cal X} + \frac{4 \pi T}{\mu} \right)^2  \right) + c.c. 
\right] \, , 
\end{aligned}
\ee
where $\mathcal{X}$ is now an ordinary chiral superfield, $Z$ and $\Psi$ are Lagrange multiplier chiral superfields and $U$ is a real superfield.
The role of $Z$ is to impose the condition that ${\cal X}$ is a three-form chiral superfield, 
while the role of $\Psi$ is to make it nilpotent as in \eqref{ShiftedX}. 
Indeed, varying \eqref{SG-EQUIV} with respect to $Z$ one gets \eqref{sugra-X}. 
Now we want instead to derive the on-shell equivalent form with the standard de Sitter supergravity 
because it is the only way to identify correctly the supersymmetry breaking scale, 
i.e. the scale that suppresses the goldstino interactions and so gives the cut-off. 
First we integrate out $U$ and we get 
\be
Z = i \tilde f \,, 
\ee
where $\tilde f$ is a positive real  constant. 
Then since ${\cal X}$ is still a standard chiral superfield, we can redefine it as 
\be
{\cal X} = X - \frac{4 \pi T}{\mu} \,, 
\ee
which brings the superspace Lagrangian to the form 
\be
\label{standard dS}
{\cal L} = - 3 \int d^4 \theta E \, e^{- \frac13 |X|^2}  
+ \left[ 
\int d^2 \Theta \, 2 {\cal E} \left( m_{3/2} - \frac{4 \pi i \tilde f T}{\mu} + i \tilde f  X  + \Psi X^2  \right) + c.c. 
\right] \,. 
\ee
Now we integrate out $\Psi$ and we have a standard de Sitter supergravity with $X^2=0$ and superpotential 
\be
\label{on-shell-equiv}
W = m_{3/2} - \frac{4 \pi i \tilde f  T}{\mu} + i \tilde f  X \, . 
\ee
The supersymmetry breaking order parameter is by definition (in standard de Sitter supergravity) 
the modulus of the parameter in front of $X$ in the superpotental \cite{Bergshoeff:2015tra}. 
Thus we see that the supersymmetry breaking order parameter is given by $\tilde f$, 
which means the cut-off is 
\be
\Lambda_{cut-off} = \sqrt{\tilde{f}} \, . 
\ee 
Note also that the gravitino mass is now complex and it is not given only by $m_{3/2}$. 
Finally, the scalar potential is 
\be
\label{V-on-equiv}
V = \tilde f^2 - 3 m_{3/2}^2 - 3 \left(\frac{4 \pi \tilde f  T}{\mu} \right)^2 \, . 
\ee
Now we have to relate $\tilde f$ to the flux $n$ of our component form discussion. 
Since the two systems are equivalent, 
the simplest way to do this is to relate them by asking that in both cases the scalar potentials match. 
We thus identify $V$ in \eqref{V-on-equiv} 
with the scalar potential in \eqref{boson-3forms} and we have 
\be
\tilde f = {\rm n} \left( 1 - 3 \left( \frac{4 \pi T}{\mu} \right)^2 \right)^{-1/2} \, . 
\ee 
With this identification we now have an exact match between the standard de Sitter supergravity and the three-form de Sitter supergravity, 
and we have also identified correctly the cut-off.

Since we wish to study decay from very shallow dS to AdS in one jump, 
this means that 
\be
{\rm n}_i = 0 \ \to \ {\rm n}_o = \Delta {\rm n} = Q/2 \, . 
\ee
As a result, outside of the membrane we have for the supersymmetry breaking scale and for the cut-off 
(using also the expression for $\mu$ given by \eqref{muQT}) 
\be
\label{theCO}
\Lambda_{cut-off} =\sqrt{\tilde{f}_o} = \sqrt{\frac{Q}{2}} \left( 1 + \frac{3T^2}{Q^2} \right)^{1/4} \,. 
\ee
In the mean time, the membrane tension is generically expected to be of order $|T| \geq \Lambda_{cut-off}^3$ \cite{Bedroya:2020rac}. 
Taking into account that $Q\ll1$ and $|T|\ll1$ the simplest way to have a self-consistent setup is to assume 
overcharged 2-branes. This is in accordance with the WGC \cite{ArkaniHamed:2006dz}, which states that 
for a theory with p-forms coupled to gravity there should exist an electrically charged state of charge $Q$ and tension $T$ which satisfies
\begin{eqnarray}
T_3 \leq |Q|. \label{TTQ}
\end{eqnarray} 
In particular, we will assume that there exists
a super-extremal 2-brane with charge much above its  tension 
\be
T_3 \ll Q \, ,  \label{TQQ}
\ee
keeping always in mind that we are in Planck units. This  condition leads to a SUSY breaking 
scale $f$  much below the Planck scale. The closer we are to the saturation of (\ref{TTQ}), the higher the cutoff as we will see below. For the moment we will take our super-extremal 2-branes to satisfy (\ref{TQQ}) in order to have a reasonable scale for SUSY breaking.
Then \eqref{theCO} gives 
\be 
\label{TQrel}
|T| \simeq \left(\frac{Q}{2}\right)^{3/2} \ll Q \ll 1 \,, 
\ee
which also means $|T| \ll |\mu|$.

Now we are ready to check the decay time of the de Sitter supergravity. 
We will directly follow \cite{BT1,BT2,deAlwis:2013gka,Giudice:2019iwl} taking into account 
that the tunneling probability per unit time per unit volume is 
\be
P \sim e^{- B} \, , 
\ee
and evaluate $B$, which is the difference between the on-shell value of the instanton action 
describing the membrane nucleation and the background action. 
We are interested in particular in late-time cosmological models which have $\Lambda_o \sim 10^{-120}$, 
which means that for our purposes it is sufficient to study the simple setup where 
\be
\Lambda_o \simeq 0 \,. 
\ee
Then taking into account \eqref{TQrel} we have 
\be
\label{EpsilonT}
\epsilon = \Lambda_o - \Lambda_i = \frac14 Q^2 \gg T^2 \ , \quad T^4 \simeq \epsilon^3  \, , 
\ee 
which following closely the analysis in \cite{deAlwis:2013gka} give independent of the details a decay rate of order 
\be
\label{B-DS}
B = \frac{27 \pi^2 T^4}{2 \epsilon^3} \frac{1}{[1 - 3 T^2 / 4 \epsilon]^2} \simeq \frac{27 \pi^2}{2} \, . 
\ee
We can check the self-consistency of our analysis by inserting realistic values for the scales involved. 
First of all a sufficient condition to avoid decay within the life-time of our universe is $B \gtrsim 10^3$, see e.g. \cite{Bousso:2000xa}. 
If we ask for the late-time de Sitter phase to have a gravitino mass of order 
\be
\label{TESTm}
m_{3/2} \sim 10^{-6} \, , 
\ee
then via the cut-off the membrane charge and tension take the values 
\be
Q \sim 10^{-6} \ , \quad |T| \sim 10^{-9} \, . 
\ee
Notice that the modulus of the effective value of the gravitino mass dictated by \eqref{on-shell-equiv}, 
that is $|m_{3/2} - 4 \pi i \tilde f  T / \mu|$, 
is essentially controlled by $m_{3/2}$ consistently with \eqref{TESTm}. 
Then the difference between the swallow dS and the AdS vacuum is of order 
\be
\epsilon \sim 10^{-12} \, , 
\ee
which leads to a value for $B<10^3$ identical to the one we found 
from the general analysis.

Even though we have until now asked that $|T|\ll Q$, 
this is not a necessary condition rather it is sufficient. 
Indeed, 
we can relax this hierarchy as long as we ask that $|T|<Q$ with $Q$ having of course realistic values. 
Realistic values for $Q$ means that it definitely cannot exceed a unit value, 
and that it leads to a cut-off safely below the Planck scale. 
We can then have an explicit example for clarity 
\be
Q = 10^{-2} \ , \quad |T| = 2^{-3/2} 10^{-3} \,. 
\ee
From our previous analysis such a value for $Q$ readily gives 
\be 
\Lambda_{cut-off} = 2^{-1/2} 10^{-1} \,,  
\ee
so one could not have a larger $Q$ and remain safely within supergravity. 
For this setup however, 
the condition \eqref{EpsilonT} still holds and so we still find the decay to be given by \eqref{B-DS}, 
giving again a short-lived de Sitter.

Finally let us see what happens after the decay once the nilpotent superfield evaporates. 
To verify the vacuum structure we can simply study the equivalent standard supergravity Lagrangian \eqref{standard dS}, 
with $\tilde f \equiv 0$ and $X \equiv 0$. 
Such setup gives the standard supersymmetric anti de Sitter supergravity with vacuum energy $V = -3 m_{3/2}^2$. However, 
the full dynamics of the theory are controlled by the original three-form action \eqref{TOT-ACTION} with ${\cal X} \equiv - 4  \pi T / \mu$, 
and will describe the pure three-form supergravity coupled to a supermembrane as studied in \cite{Ovrut:1997ur} 
but now in the presence of a superspace cosmological constant given by the superpotential $W = m_{3/2}$. 
The reason ${\cal X} \equiv - 4  \pi T / \mu$ forces the structure of the supergravity multiplet to change is because for the 
$F$ component of $\mathcal{X}$ we will have $F\equiv0$, which can only be solved by setting, apart from ${\cal F}=0$, 
also ${\rm Im} M = - (\mu D^a C_a) / (4 \pi T)$ 
up to fermions, 
as can be seen from \eqref{F-bosonic-nilpotent} 
- this procedure of exchanging ${\rm Im} M$ with a three-form is described in \cite{Ovrut:1997ur}. 
If no further membrane nucleation takes place the vacuum energy will remain $V = -3 m_{3/2}^2$. 
This is in agreement with our discussion in supersymmetry where we see that a membrane still remains coupled only 
with the background once the goldstino evaporates. 
In supergravity however, 
the only membrane that can couple to the pure AdS supergravity background is the one given in \cite{Ovrut:1997ur} 
which at the same time forces the structure of the supergravity multiplet to have a three-form auxiliary field.

\subsection{Decay without complete evaporation}

\begin{figure}[ht]
\centering 
  \includegraphics[scale=0.9]{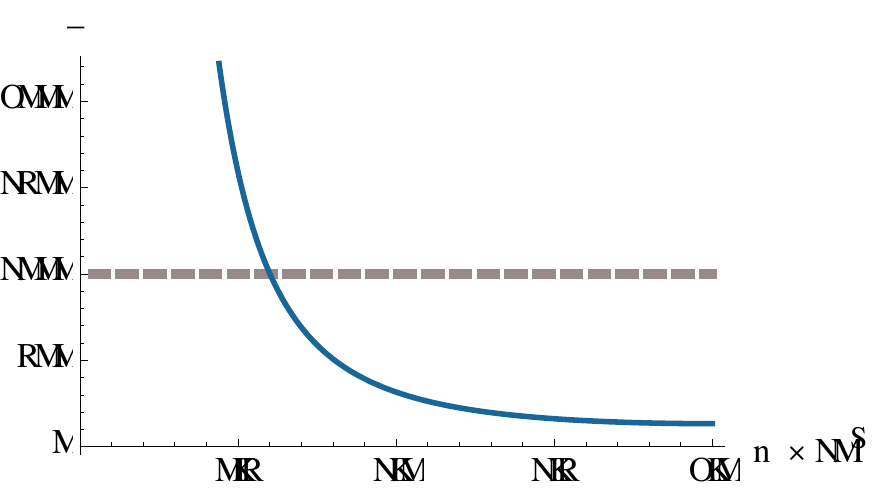} 
\caption{ 
{\it The drop of $B$ as $Q$ increases for the working example of equation \eqref{WEX}. 
We see that a complete evaporation of the goldstino is favored because $Q$ reaches a maximum value, which gives ${\rm n}_i=0 $,  
while $B$ is minimized. 
The dashed line represents a crude estimate for the minimum value of $B$ such that the vacuum has 
safely a life-time much larger than our Universe.} 
\label{plot-QVSB}} 
\end{figure}

Until now we have assumed that the goldstino completely evaporates in one go. 
Here we will elaborate on what happens if the goldstino does not directly evaporate in one go, 
i.e. if there is some residual flux left after the tunneling 
\be 
{\rm n}_i = {\rm n}_o - \frac{1}{2} Q \ne 0 \, . 
\ee 
We will study this situation in a generic scenario. 
Let us then have as a working example before the first nucleation 
\be
\label{WEX}
\Lambda_o = 10^{-120}  \ , \quad {\rm n}_o = 10^{-6} = \Lambda_{cut-off}^2 \ , \quad |T| = \Lambda_{cut-off}^3 = 10^{-9} \, . 
\ee
We want to get an estimate of the decay so we will not fix $Q$, 
rather we will give a figure of the expected decay probability in terms of $Q$. 
The vacuum energy after the first nucleation will be 
\be
\Lambda_i = \Lambda_o - \frac{1}{2} Q \left(2 {\rm n}_o - \frac{1}{2} Q\right) \,. 
\ee
The decay rate is depicted in figure \ref{plot-QVSB} where we see that a full evaporation of the nilpotent goldstino multiplet is favored. 
In the latter case we have $\epsilon^3 \simeq T^4$ which gives $B \simeq 27 \pi^2 /2$ independent of the 
details. 

\section{Decay in KKLT}

The nilpotent superfield description of the KKLT setup is given for example in \cite{Ferrara:2014kva,Bergshoeff:2015jxa} 
and has an extra chiral multiplet $\Phi$ on top of the nilpotent one. 
One works as before with an action of the form \eqref{TOT-ACTION} 
with the super-membrane sector given by \eqref{MEMG-SG} and with the appropriate boundary terms. 
Taking into account the shifted nilpotency condition \eqref{ShiftedX} 
and the discussion in the previous section 
one can easily guess that the forms of the K\"ahler potential and superpotential for the full three-form formulation of KKLT are 
$K = -3 \, log \left( \Phi + \overline \Phi - | {\cal X} + 4 \pi T / \mu |^2/3 \right)$ 
and $W= \tilde W_0 + \alpha \, e^{- a \Phi} $, 
where $\tilde W_0$, $a$ and $\alpha$ are real constants. 
As before we need the shifted nilpotency condition \eqref{ShiftedX} 
because the vacuum expectation value of ${\cal X}|$ sustains the membrane tension. 
We do not need to discuss again in detail the properties of the three-form formulation of the action, 
as it is essentially covered by the previous section. 
However, we need to identify the effective membrane charge, 
which dictates the flux jump. 
The relevant bosonic sector after the Weyl rescaling is 
\be
{\cal L}_{DC} = e \left( e^{-K/3} D^a C_a - e^{K/3} \frac{4 \pi T}{\mu} {\rm Im} W_\phi \right)^2 
+ \text{boundary terms} \, , 
\ee 
whereas the WZ term is $- (\mu / 4 \pi) \int_{\mathcal{M}_3} {\cal C}_3 $. 
Once we integrate out $C_3$ we find (taking into account the boundary terms) 
\be
\label{LDCDC}
D^a C_a = e^{K/3} {\rm n} + e^{2K/3} \frac{4 \pi T}{\mu} {\rm Im} W_\phi 
\quad \to \quad   {\cal L}_{DC} = - e \, {\rm n}^2 - e e^{K/3} \frac{8 \pi T}{\mu}{\rm Im} W_\phi \, {\rm n} \, , 
\ee
where ${\rm n} = e^{K/3}   \times \text{constant} $ and the effective membrane charge 
(the 2-brane charge under the canonically normalized 3-form) 
that dictates the flux jump is 
\be
\label{KKLT-eff-Q}
\frac{Q}{2} = - \frac{\mu e^{K/3}}{8 \pi} = \Delta {\rm n} \ , \quad \mu < 0 \, . 
\ee 
Supersymmetry is restored when $n=0$. 
Of course the full vacuum energy will receive also other contributions on top of the flux-related part \eqref{LDCDC}.

Note that the flux contribution to the vacuum energy in \eqref{LDCDC} is not only $-{\rm n}^2$ but there 
is also a term linear in the flux, n, and in Im$W_\phi$. 
The importance of the latter term depends on the details of the model. 
For our analysis here we are only interested in its contribution when we are discussing a realistic KKLT setup. 
In such a scenario one can find that 
\be
e^{K/3} \frac{8 \pi T}{\mu}{\rm Im} W_\phi \, {\rm n} \ 
\simeq \ 
{\rm n}^2 \, \frac{T^2 a \alpha}{\mu^2 \tilde W_0} e^{- a {\rm Re}[\phi_{dS}]} \,, 
\ee
where $\phi_{dS}$ is the value of $\phi$ on the de Sitter vacuum. 
Therefore, 
in order to have the flux contribution to the vacuum energy to be dominated by the n$^2$ term in \eqref{LDCDC}, 
we have to satisfy 
\be
\label{COND-EXT}
\frac{T^2 a \alpha}{\mu^2 \tilde W_0} e^{- a {\rm Re}[\phi_{dS}]} \ll 1\,. 
\ee
Indeed in a realistic KKLT setup it holds that the condition \eqref{COND-EXT} 
is satisfied because the expression on the left hand side is of order $10^{-9}$. 
We will in fact explicitly verify this with a specific example. 
Therefore we will from now on assume that 
\be
\label{LDCDC-IMP}
{\cal L}_{DC} = - e \, {\rm n}^2 \, , 
\ee
and we will a posteriori verify that the consistency requirements for such condition do hold.

Now we can proceed directly to the 
equivalent description of the bulk sector in the form of standard supergravity which will allow us to derive 
both the cut-off and the difference in the vacuum energy before and after the goldstino evaporation, 
i.e. the ingredients that dictate the decay rate. 
We will consider the effective theory that is appropriate for the strong warping regime \cite{Ferrara:2014kva,Bergshoeff:2015jxa}. 
The equivalent bulk sector we need to work with is given by the superspace Lagrangian 
\be
\label{SG-EQUIV-KKLT1}
\begin{aligned}
{\cal L} = & \int d^4 \theta E \left( -3 \Phi -3 \overline \Phi + \Big{|} {\cal X} + \frac{4 \pi T}{\mu} \Big{|}^2 
- U (Z + \overline Z) \right) 
\\
& + \left[ 
\int d^2 \Theta \, 2 {\cal E} \left( \tilde W_0 + \alpha \, e^{- a \Phi} 
+ Z {\cal X} + \Psi \left( {\cal X} + \frac{4 \pi T}{\mu} \right)^2  \right) + c.c. 
\right] \, ,
\end{aligned}
\ee 
where $\mathcal{X}$ is now an ordinary chiral superfield, $Z$ and $\Psi$ are Lagrange multiplier chiral superfields and $U$ is a real superfield. The role of $Z$ is again to impose the condition that ${\cal X}$ is a three-form chiral superfield given by \eqref{sugra-X}, 
while the role of $\Psi$ is, as usual, to make it nilpotent as in \eqref{ShiftedX}; 
integrating out $Z$ and $\mathcal{X}$ takes us to the aforementioned nilpotent three-form formulation. 
Now instead we integrate out $U$, which gives $Z = i f$, where $f$ is a real constant. 
Then, as before, we can redefine ${\cal X} = X - 4 \pi T / \mu$, which brings us to a superspace Lagrangian of the form 
\be
\label{SG-EQUIV-KKLT-2}
\begin{aligned}
{\cal L} = & \int d^4 \theta E \left( -3 (\Phi + \overline \Phi) + |X|^2 \right) 
\\
& + \left[ 
\int d^2 \Theta \, 2 {\cal E} \left( \tilde W_0 - \frac{4 \pi i f T}{\mu} 
+ \alpha \, e^{- a \Phi} 
+ i f X + \Psi X^2  \right) + c.c. 
\right] \,. 
\end{aligned}
\ee
To bring this Lagrangian to the standard form presented in \cite{Ferrara:2014kva,Bergshoeff:2015jxa} we first define 
\be
\tilde W_0 - \frac{4 \pi i f T}{\mu}  \equiv W_0 e^{i r} \, , 
\ee
where both $W_0$ and $r$ are real constants. 
Note that when $|4 \pi f T| \ll |\mu \tilde W_0|$ then $|W_0| \simeq |\tilde W_0|$. 
We will ask that the latter condition is satisfied such that the value of $W_0$ does not change significantly when $f=0$. 
This is easily achieved by asking 
\be
\label{TfWm}
|T| \ll |\mu| \ , \quad |f| \sim |\tilde W_0| \sim |W_0| \,. 
\ee
Then we perform the superfield redefinitions
\be
\label{REDEFS}
X \to -i X e^{i r} \ , \quad \Phi \to \Phi - i r / a \ , \quad \Psi \to - \Psi e^{-i r} \,, 
\ee
which leave the K\"ahler potential invariant, 
accompanied by an appropriate super-Weyl transformation that acts on the 
chiral density as $2 {\cal E} \to 2 {\cal E} e^{-ir}$ but leaves the real density $E$ invariant. 
Note also that \eqref{REDEFS} changes only the imaginary part of the lowest component of $\Phi$. 
After these redefinitions, we integrate out $\Psi$ and recover the standard nilpotency condition $X^2=0$. 
This brings us to the desired form of the bulk Lagrangian, 
which is standard supergravity with K\"ahler potential and superpotential \cite{Ferrara:2014kva,Bergshoeff:2015jxa}
\be
K = -3 \, log \left( \Phi + \overline \Phi - \frac13 X \overline X \right) 
\ , \quad 
W = W_0 + \alpha \, e^{- a \Phi} + f X \, . 
\ee
Now, 
what the three-form formulation buys us is that we can have membrane nucleations that give 
\be
f_{out} \ne 0 \ \xrightarrow{decay} \ f_{in} = 0 \, . 
\ee 
When $f \ne 0$ we have supersymmetry spontaneously broken and intrinsically non-linearly realized 
as in KKLT \cite{Ferrara:2014kva,Bergshoeff:2015jxa}, 
whereas 
when $f=0$ we have a supersymmetric AdS coupled to the chiral multiplet $\Phi$, 
while the goldstino multiplet evaporates. 
We repeat that such a transition can not happen in a continuous way, 
and it is only possible here because we have a discrete jump in the value of $f$ induced by the membrane. 
The scalar potential has the typical 4D N=1 form  
\be
V = e^{K} \left( g^{i \overline j} D_i W \overline D_{\overline{j}} \overline W  - 3 |W|^2 \right) \, , 
\ee
where $D_i W = W_i + K_i W$ and $\phi^i$ refers to $\phi=\Phi|$ and $A=X|$.

\begin{figure}[ht]
\centering 
  \includegraphics[scale=0.9]{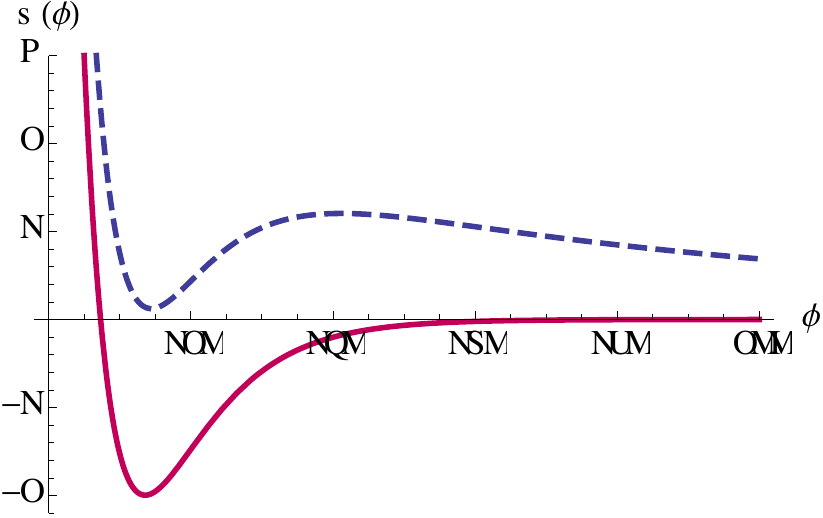} 
\caption{ 
{\it The KKLT scalar potential in units $10^{15} M_P^4$ 
(for the values $\alpha = 1$, $a = 0.1$, $W_0 = - 10^{-4}$, $f =1.05 \times 10^{-5}$) 
before and after the evaporation of $X$. Here ${\rm Im} \phi=0$. 
These values are typical for KKLT, see e.g. \cite{Kallosh:2019axr}.} 
\label{plot-KKLT}} 
\end{figure}

We are again interested in the complete evaporation of the $X$ multiplet in one go. 
We assume that when the system passes through the bubble the scalar $\phi$ keeps its position $\phi_{dS}$ 
and then it rolls to the supersymmetric position $\phi_{AdS}$. 
Our discussion essentially follows closely the pure de Sitter supergravity discussion 
because one can recast the scalar potential to the form 
\be
V = f_o^2 - \lambda \, , 
\ee
where we have the effective supersymmetry breaking scale and a negative term that remains almost unchanged during the transition 
\be
f_o^2 = e^K |D_A W|^2 \Big{|}_{\phi_{dS}}  \ , \quad - \lambda = e^K ( |D_\phi W|^2 - 3 |W|^2) \Big{|}_{\phi_{dS}} \, , 
\ee
where we use the standard notation $|D_\phi W|^2 \equiv (K_{\phi \overline{\phi}})^{-1} D_\phi W \overline D_{\overline \phi} \overline W$ and 
similarly for $|D_A W|^2$ because the K\"ahler metric is diagonal. 
From \cite{Hasegawa:2015} we can see that the cut-off of such a  theory is 
\be
\Lambda_{cut-off} = \sqrt{f_o} = e^{K/6} \sqrt f \,, 
\ee 
where we have taken into account that $2 \phi_{dS} = e^{-K/3}$, which defines the effective membrane tension to be 
\be
\label{KKLT-TEFF}
T_3 = \Lambda_{cut-off}^3 = e^{K/2} f^{3/2} \, . 
\ee
Note that this is not given by $T$ because of the Weyl rescaling. 
Indeed from the Weyl rescaling of the Nambu--Goto term we find $T_3 = |T| e^{K/2}$ 
which in the end gives $|T|$ in terms of $f$ to be 
\be 
\label{TFK}
|T| = f^{3/2} \, . 
\ee
Then for the energy difference between the ``out'' and ``in'' backgrounds 
taking into account \eqref{TfWm} we have 
\be
\label{E-KKLT}
\epsilon = V_o - V_i \simeq f_o^2 = e^{2K/3} f^2 \,. 
\ee

Here we can perform a series of cross-checks for our various assumptions. 
First, we perform a fast cross-check of the self-consistency of the assumption \eqref{TfWm}. 
From the fact that the energy difference is essentially due to the jump in the flux value 
(from \eqref{LDCDC-IMP} in fact the difference is $(\Delta {\rm n})^2$ because the flux is neutralized in one jump), 
which is always $\Delta {\rm n} = Q/2$, 
we have from \eqref{E-KKLT} that 
\be 
\label{QFK}
f \simeq e^{-K/3} Q/2 \, . 
\ee
Then inserting \eqref{KKLT-eff-Q} into \eqref{QFK} 
we find $f \simeq -\mu / 8 \pi$ 
and then inserting the latter into \eqref{TFK} 
we find a self-consistency relation between $|T|$ and $\mu$ of the form 
\be
|T| \simeq \left( \frac{|\mu|}{8 \pi} \right)^{3/2}\,. 
\ee 
Taking into account that $|\mu| \ll 1 $ we see a complete agreement with $|T| \ll |\mu|$ 
and thus do not jeopardize \eqref{TfWm}. 
Meanwhile note that \eqref{TQrel} now holds for the effective membrane tension and charge, 
that is $T_3$ and $Q$ respectively, 
therefore we are again working with a super-extremal 2-brane. 
The final cross check concerns the validity of \eqref{COND-EXT}. 
We have already seen that $T^2 \ll \mu^2$, which is in support of \eqref{COND-EXT}, 
however, 
one can be worried that the contribution from $\tilde W_0^{-1} a \alpha e^{- a {\rm Re}[\phi_{dS}]}$ 
may be larger than unit. 
The simplest way to check that $|\tilde W_0^{-1} a \alpha e^{- a {\rm Re}[\phi_{dS}]}| < 1$ 
is to use the values we have in the working example of figure \ref{plot-KKLT}, 
taking into account that $\tilde W_0 \simeq W_0$ due to \eqref{TfWm}. 
Indeed, 
for the values of the working example of figure 
\ref{plot-KKLT} we have $|\tilde W_0^{-1} a \alpha e^{- a {\rm Re}[\phi_{dS}]}| = {\cal O}(10^{-2})$, 
and one can easily verify that the expression on the left hand side of \eqref{COND-EXT} is of order $10^{-9}$.

We can close this section by evaluating the decay rate. 
From \eqref{KKLT-TEFF} and \eqref{E-KKLT} we see 
that $T_3^4 \simeq \epsilon^3$ and thus for $V_o \simeq 0$ (as in a realistic KKLT scenario) 
the function $B$ that controls the decay rate takes the form 
\be
B \simeq \frac{27 \pi^2}{2} \, . 
\ee
We conclude that we have again an insufficient suppression of the decay channel also in KKLT. 
Essentially as far as the decay is concerned the situation in KKLT is analogous to the pure de Sitter supergravity, 
as both describe short-lived dS vacua.

\section{Outlook}

We have presented a system that interpolates between 
a state with intrinsically non-linearly realized supersymmetry 
and a supersymmetric ground state. 
We have shown that there is a way to describe the process of goldstino superfield evaporation which 
we have argued is essential for such transitions. 
We have investigated the implications of such decay channels and found that they pose a threat for supergravity de Sitter model building, 
in agreement with the recent developments \cite{Danielsson:2018ztv,Obied:2018sgi}\footnote{The no-dS conjectures have been 
discussed in N=1 supergravity explicitly in \cite{Ferrara:2019tmu} from the perspective of restricting the scalar potential.}. 
The decay time depends of course on the membrane charge, 
but if the transition to AdS (and so the goldstino evaporation) happens in one go, 
then the life-time of the de Sitter is found to be unrealistically small. 
In particular such imminent decay of de Sitter requires $T_{membrane} \ll Q_{membrane}$, which we do not have a way to justify at this 
point from a pure supergravity perspective and deserves a careful further investigation; 
maybe it is possible to find EFT/swampland arguments in favor or against such 
property.\footnote{Similarly, 
we do not have arguments for why the membrane tension cannot be larger than $\Lambda_{cut-off}^3$ (which would make the KKLT vacuum long-lived), and we leave a careful investigation for future work. 
Note that in particular for the evaluation of the decay time and the membrane tension one has to take into account 
more recent developments as for example \cite{Danielsson:2014yga}. } 
The only support that we have in favor of this hierarchy now is that it is dictated by asking that 
the goldstino evaporates in one single transition, 
which from the anti-D3-brane perspective \cite{Ferrara:2014kva,Bergshoeff:2015jxa,Akrami:2018ylq} 
would mean that a single anti-D3-brane, as in KKLT (see in particular \cite{Polchinski:2015bea,Akrami:2018ylq}), 
decays of course in one go.\footnote{In KPV in any case the decay happens in one go independent of the number of anti-branes.}

Our analysis motivates the further use of the nilpotent goldstino three-forms  
and the associated super 2-branes as a way to implement the various swampland 
conjectures within an effective 4D N=1 supergravity framework. 
Note that here we worked with a goldstino that is embedded within a constrained three-form chiral superfield, 
but one can perform a similar investigation for a goldstino living inside a constrained three-form vector multiplet 
within the lines of \cite{Benakli:2017yar,Cribiori:2017laj,Cribiori:2018jjh,Cribiori:2020wch}. 
Another direction is to study the goldstino decay in the presence of double three-form multiplets \cite{Kuzenko:2017vil,Bandos:2018gjp}. 
Furthermore, 
it is worth performing a similar analysis but focused on the inflationary 
cosmological phase instead of the late-time as we did here.

Finally, the decay channel we described here should be perceived 
as an effective description of a more fundamental underlying transition 
that may become smooth and more complicated when we turn to full string theory. 
This procedure may eventually resemble some type of brane-flux 
annihilation \cite{Kachru:2002gs,Frey:2003dm,Gautason:2015tla,Garcia-Etxebarria:2015lif} 
but we leave such investigation for a future work.

\section*{Acknowledgements}

We thank Thomas Van Riet for discussion. 
The work of FF is supported by the STARS-StG Grant ``SUGRA-MAX''.

\end{document}